\documentclass[12pt]{article}
\usepackage{graphicx}
\usepackage{psfrag}
\usepackage{color}
\usepackage{subfigure}
\usepackage{amsmath}
\usepackage{longtable}
\newcommand{\rf}[1]{(\ref{#1})}
\newcommand{\sh}{\sinh}
\newcommand{\ch}{\cosh}
\newcommand{\half}{\frac{1}{2}}
\begin{document}

\pagestyle{empty}
\begin{titlepage}
\vspace*{2cm}
\begin{center}
\Large
\vspace{.2cm}
{\bf \mbox{Critical Properties of an Integrable }\\}
\vspace{.2cm}
{\bf \mbox{Supersymmetric Electronic Model}\\}
\vspace{1cm}
\normalsize
\mbox{A.L. Malvezzi}\\
\vspace{0.2cm}
\mbox{Universidade Estadual Paulista-UNESP, Faculdade de Ci\^encias}\\
\mbox{Departamento de F\'\i sica, CP 473, 17033-360, Bauru, S.P., Brazil}\\
\vspace{.5cm}
\mbox{M.J. Martins}\\
\mbox{Universidade Federal de S\~ao Carlos, Departamento de F\'isica}\\
\mbox{CP 676, 13565-905, S\~ao Carlos, S.P., Brazil}\\
\vspace{0.2cm}
\vspace*{2.cm}
%{\bf \mbox{Abstract}}\\
%\vspace{.8cm}
\begin{abstract}
We investigate the physical properties of an integrable extension of
the Hubbard model with a free parameter $\gamma$ related to the quantum
deformation of the superalgebra $sl(2|2)^{(2)}$. The Bethe ansatz solution
is used to determine the nature of the spin and charge excitations.
The dispersion
relation of the charge branch is given by a peculiar product between
energy-momenta functions exhibiting massless and massive behaviors. 
The study of the finite-size corrections to the spectrum reveals us that
the underlying conformal theory has central charge $c=-1$ 
and critical exponents depending on the parameter $\gamma$. We note
that exact results at the isotropic point $\gamma=0$ can be established
without recourse to the Bethe ansatz solution. 
\end{abstract}
\vspace{0.5cm}
PACS numbers: 75.10.Jm, 71.10.Pm, 64.60.F-, 64.60.an
\end{center}
\end{titlepage}
\newpage

\pagestyle{plain}
\pagenumbering{arabic}
\section{Introduction}
\label{sec:introd}

The study of electron correlation effects in one-dimensional systems have by now
attracted the attention of theorists for more than a half-century. The physical
behavior of one-dimensional correlated electron models are expected to be
drastically different from that of free electrons \cite{HA}. It turns out that the
basic excitations
have a collective character and non-perturbative techniques becomes essential.
In this context, electronic lattice models solvable by Bethe ansatz have
provided relevant insights into the physical properties of such systems \cite{LI,SU,ST}.
Of particular interest are integrable extensions of Hubbard model derived exploring solutions
of the Yang-Baxter equation with two fermionic and two bosonic degrees of freedom \cite{KUL}.
Representative examples are the models associated with the four dimensional representations
of the $sl(2|2)$ and $gl(2|1)$ Lie superalgebras \cite{EKS,BRA,BER}. We remark that 
generalizations of the Hubbard models based on the quantum deformations of such algebras \cite{KU,JL,MR}
as well as on the central extension of $sl(2|2)$ \cite{AL} have also been discussed in
the literature.

The purpose of this paper is to investigate the critical properties of an extended
Hubbard model based on the quantum deformation of the twisted $sl(2|2)^{(2)}$ algebra \cite{JL}. 
We recall here that this model appears to provide a lattice regularization of an interesting 
integrable (1+1)-dimensional
quantum field of two coupled massive Dirac fermions \cite{SA1}.
Though
the respective Bethe ansatz solution is known \cite{MR} it has not yet been explored to
extract information about the physical properties of such lattice electronic model. 
Following \cite{MR} the model
Hamiltonian can be re-written as,
\begin{eqnarray}\label{H}
H &=& \sum_{i=1}^{L} \sum_{\sigma=\pm} \left [ c_{i, \sigma}^{\dagger}
c_{i+1, \sigma} +
h.c.  \right ] \left [ 1- X_{\sigma}n_{i, -\sigma}
-{\bar X}_{\sigma}n_{i+1, -\sigma} \right ]
+ U\sum_{i=1}^{L}  n_{i,+} n_{i,-} 
\nonumber \\
&+& V\sum_{i=1}^{L}
\left[n_{i,+} n_{i+1,-} +n_{i,-}n_{i+1,+} \right]
+ Y\sum_{i=1}^{L} \left [ c_{i,+}^{\dagger} c_{i,-}^{\dagger}
c_{i+1,-} c_{i+1,+}+h.c. \right ]
\nonumber \\
&+&J\sum_{i=1}^{L} \left[ c_{i,+}^{\dagger} c_{i+1,-}^{\dagger} c_{i,-} c_{i+1,+} +h.c \right ]
-\mu \sum_{i=1}^{L}\left[ n_{i,+}+n_{i,-} \right ]
\end{eqnarray}
where $c_{i, \sigma}^{\dagger}$ and $c_{i, \sigma}$ are fermionic creation and annihilation
operators with spin index $\sigma=\pm$ acting on a chain of length $L$. The operator
$n_{i,\sigma}=c_{i, \sigma}^{\dagger}c_{i, \sigma}$ represents the number of electrons
with spin $\sigma$ on the $i$-th site.

Apart from the standard kinetic hopping amplitude and the on-site Coulomb term
$U$ we see that Hamiltonian \rf{H} contains additional interaction terms. They are
the bond-charge hopping amplitudes $X_{\sigma}$ and ${\bar X}_{\sigma}$, the Coulomb
interaction $V$ among electrons at nearest-neighbor sites, the spin-spin exchange term $J$,
the pair-hopping amplitude $Y$ besides the chemical potential amplitude $\mu$.
Integrability constraints the couplings of the model on the following one-parameter
manifold,
\begin{equation}\label{const1} % obs.: "const" stands for "constraints"
X_{\sigma}=1 +\sigma \sin(\gamma),  ~~ \bar{X}_{\sigma}=1 -\sigma \sin(\gamma),~~ 
 \frac{U}{2}=V=J=Y=\cos(\gamma)  
\end{equation}
where the anisotropy $\gamma$ is related to the $q$-deformation of $sl(2|2)^{(2)}$ by $q=\exp[{i\gamma}]$.

The potential $\mu$ is in principle arbitrary since the model conserves the total number
of electrons with spin $\sigma=\pm$. However, the invariance of Hamiltonian \rf{H} by the
superalgebra $U_{q}[sl(2|2)^{(2)}]$ fixes a relation between $\mu$ and $\gamma$, namely \cite{JL,MR}
\begin{equation}\label{const2}
\mu = 2\cos{(\gamma)}.
\end{equation}
Considering the parameterization \rf{const1} and \rf{const2} one can relate the spectra of
Hamiltonian Eq.\rf{H} at the points $\gamma$ and $\pi-\gamma$. In fact, by performing a
combination of particle-hole $c_{i,\sigma} \to c_{i,\sigma}^\dagger$ and the parity
 $c_{i,\sigma} \to (-1)^i c_{i,\sigma}$ transformations one is able to find the following
relation,
\begin{equation}\label{transf}
H(\gamma) = -H(\pi-\gamma)
\end{equation}
Due to property \rf{transf} the analysis of the physical properties of Hamiltonian \rf{H} subjected to the
constraints (\ref{const1},\ref{const2})  can be restricted 
to the anti-ferromagnetic
interval $0 \le \gamma \le \pi/2$.
In this work we shall argue that the low-energy  behavior of this model in the regime $0 < \gamma \le \pi/2$
is that of a conformally invariant theory with central charge $c=-1$. The point $\gamma = 0$ is special since the model reduces
to the supersymmetric isotropic $sl(2|2)$ extended Hubbard model \cite{EKS}. In this case
it was argued that though the excitations are gapless the dispersion relations have a
non-relativistic branch \cite{EKS1,SA}. In fact, we found that for 
the electronic model (\ref{H}-\ref{const2}) the
speed of sound of the underlying low-lying excitations is proportional to $\sin(\gamma)$ which
vanishes in the $\gamma \to 0$ limit.

We have organized this paper as follows. In next Section we shall explore the
Bethe ansatz solution to determine the ground state and the nature of the excitations
of the electronic model (\ref{H}-\ref{const2}). A particular characteristic is that
the dispersion relation of charge excitations combines both the behavior of
massless and massive degrees of freedom. In Section 3 we study that finite-size 
properties of the spectrum of the Hamiltonian (\ref{H}-\ref{const2}) by both
analytical and numerical approach. We argue that the critical properties are
described by a critical line with central charge $c=-1$. Our conclusions are
summarized in Section 4.

\section{Thermodynamic limit}
\label{sec:thermod}

Here we will determine the ground state and the nature of the elementary excitations of the
electronic model of Section \ref{sec:introd}. These properties can be investigated by
exploring the diagonalization of Hamiltonian (\ref{H}-\ref{const2}) by the Bethe ansatz method.
It was found that the corresponding spectrum is parameterized by the following nested
Bethe equations \cite{MR},
%
%\begin{widetext}
%
\begin{equation}\label{bethe1}
\left [ \frac{\sh(\lambda_j/2-i\gamma/2)}{\sh(\lambda_j/2 +i\gamma/2)}
\right ]^{L} = \prod_{k=1}^{N_{+}} \frac{ \sh(\lambda_j - \mu_k -i\gamma)}
{ \sh(\lambda_j - \mu_k +i\gamma)},~~~ 
j=1, \cdots, N_{+}+N_{-}
\end{equation}
and
\begin{equation}\label{bethe2}
\prod_{k=1}^{N_{+}+N_{-}} \frac{ \sh(\mu_j -\lambda_k -i \gamma)}
{\sh(\mu_j -\lambda_k + i \gamma)} = - \prod_{k=1}^{N_{+}}
\frac{\sh(\mu_j -\mu_k- 2i \gamma)}
{\sh(\mu_j -\mu_k+ 2i \gamma)},~~~ 
 j=1, \cdots, N_{+}
\end{equation}
%
%\end{widetext}
%
where the integers $N_\sigma$ denote the total number of electrons with spin
$\sigma=\pm$. 

The eigenvalues $E(L,\gamma)$ of Hamiltonian (\ref{H}-\ref{const2}) 
are
given in terms of the variables $\lambda_j$ by,
\begin{equation}\label{spect}
E(L,\gamma)= \sum_{j=1}^{N_{+}+N_{-}} \frac{ 2 \sin^2(\gamma)}{\cos(\gamma)-\cosh(\lambda_j)}.
\end{equation}

\begin{figure}[h]
\begin{center}
%\vskip -0.5cm
\centerline{\includegraphics[width=12.0cm,clip,angle=270]{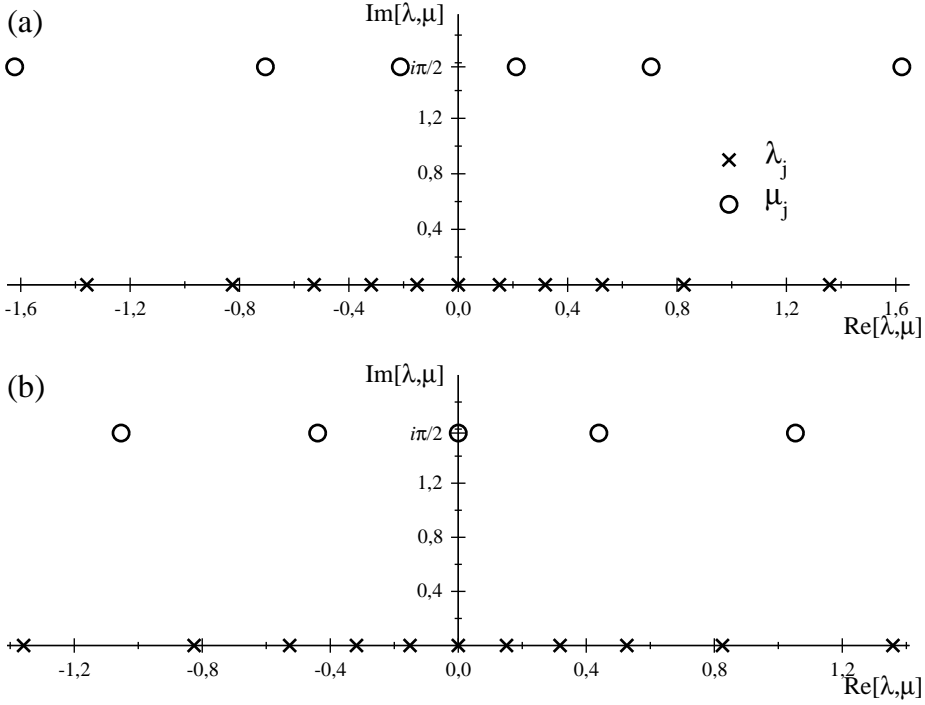}}
\caption{
The groundstate roots $\lambda_j$ (crosses) and $\mu_j$ (circles) for $\gamma = \pi/5$ and 
$L=12$ in sectors (a) $N_+ = L/2\ ,N_- = L/2 - 1$ and (b) $N_+ = L/2 - 1\ ,N_- = L/2$.
We note that the roots $\lambda_j$ are the same for both sectors.
}
\label{roots1}
%\vskip -1.0cm
\end{center}
\end{figure}

To make further progress it is important to identify the distribution of roots
$\{\lambda_j , \mu_k \}$ on the complex plane which reproduce the low-lying
energies of Hamiltonian (\ref{H}-\ref{const2}). 
This task is performed by first determining the particle number sectors of the low-lying
eigenvalues. This is done by means of brute force diagonalization of the Hamiltonian
for small chains $L \le 12$ and a few values of the parameter $\gamma$. We then compare
these eigenvalues with the results coming from the numerical analysis of the solutions
of the Bethe ansatz equations (\ref{bethe1}- \ref{spect}). By performing this analysis
we find that the
ground state in the regime $0 < \gamma \le \pi/2$ for $L$ even sits in sectors
$N_+ = L/2 \pm 1,\ N_- =  L/2$ or $N_+ = L/2,\ N_- = L/2 \pm 1$ and therefore it is four-fold
degenerated. Due to the particle-hole symmetry it is sufficient to determine the respective
pattern of the Bethe roots $\{\lambda_j , \mu_k \}$ for the sector with the minimum
possible number of roots. In Figure \ref{roots1} we exhibit the ground state Bethe roots
for $L=12$ in sectors $N_+ = L/2,\ N_- = L/2-1$ and $N_+ = L/2-1 ,\ N_- = L/2 $.
We clearly see that the roots $\lambda_j$ are real while $\mu_k$ have a fixed imaginary part
at $i\pi/2$.
\begin{figure}[h]
\begin{center}
%\vskip -0.5cm
\centerline{\includegraphics[width=12.0cm,clip,angle=270]{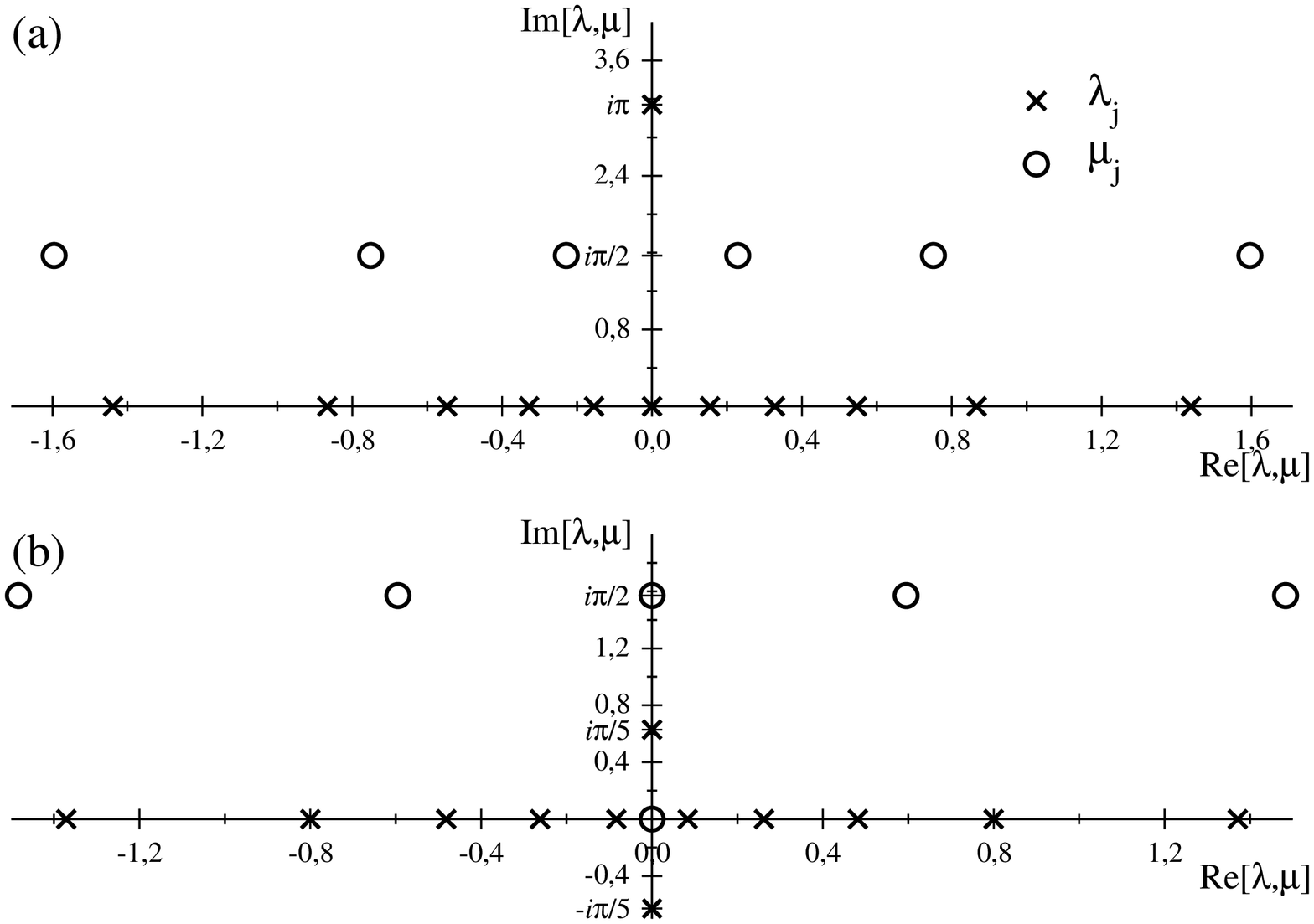}}
\caption{
The first excited state roots $\lambda_j$ (crosses) and $\mu_j$ (circles) for $\gamma = \pi/5$ and
$L=12$. Note that (b) has two roots $\lambda_j$ fixed at $\pm i\pi/5$.
}
\label{roots2}
%\vskip -1.0cm
\end{center}
\end{figure}
The first excited state is double degenerated and lies in sector $N_+ = N_- = L/2$.
In Figure \ref{roots2} we show the corresponding Bethe roots $\{\lambda_j , \mu_k \}$ for
$L=12$. 
By performing this analysis for the low-energy excitations we find that they can
be described mostly in terms of real variables when the second Bethe roots $\mu_k$ is
shifted by the complex number $i\pi/2$. Considering this discussion we find convenient
to introduce the following variables,
\begin{equation}\label{newvar}
\lambda_j=\lambda_j^{(1)},~~\mu_j = \lambda_j^{(2)} +i \frac{\pi}{2}
\end{equation}
where $\lambda_j^{(a)} \in \Re $ for $a=1,2$.

Now by substituting Eq.\rf{newvar} in the Bethe ansatz equations (\ref{bethe1},\ref{bethe2})
and afterwards by taking their logarithms we find that the resulting relations for
$\lambda_j^{(a)}$ are,
\begin{equation}\label{nbethe1}
L \Phi(\frac{\lambda_{j}^{(1)}}{2}, \frac{\gamma}{2})= 2 \pi Q_{j}^{(1)}
-\sum_{k=1}^{N_{+}} \Phi (\lambda_{j}^{(1)} - \lambda_{k}^{(2)},\frac{\pi}{2}-\gamma),~~~ 
j=1,\cdots, N_{+}+N_{-}
\end{equation}
and
\begin{equation}\label{nbethe2}
-\sum_{\stackrel{k=1}{k \neq j}}^{N_{+}} \Phi (\lambda_{j}^{(2)} - \lambda_{k}^{(2)},2 \gamma)
+2 \pi Q_{j}^{(2)}= 
\sum_{k=1}^{N_{+}+N_{-}} \Phi (\lambda_{j}^{(2)} - \lambda_{k}^{(1)},\frac{\pi}{2}-\gamma),
~~~j=1,\cdots, N_{+}
\end{equation}
where function $\Phi(\lambda,\gamma) = 2 \arctan{\left[ \cot{\left(\gamma \right)}\tanh{\left(\lambda \right)} \right]}$.

The numbers $Q_j^{(a)}$ define the many possible logarithm branches and in general are integers or
half-integers. Considering our previous numerical analysis we find that the low-lying spectrum
is well described by the following sequence of $Q_j^{(a)}$ numbers,
\begin{eqnarray}\label{Qs}
Q_{j}^{(1)}&=& -\frac{1}{2} \left[ L-n_{+}-n_{-} -1 \right] + j - 1,~~~ 
j=1,\dots, L-n_{+}-n_{-}  \\
Q_{j}^{(2)}&=& -\frac{1}{2} \left[ \frac{L}{2}-n_{+} -1 \right] + j - 1,~~~ 
j=1,\dots ,\frac{L}{2} - n_{+}
\end{eqnarray}
where $n_\pm$ are integers labeling the sector with $N_\pm = L/2 - n_\pm$ particles with spin $\sigma = \pm$.

For large $L$ the number of roots tend towards a continuous distribution on the real axis whose density
can be defined in terms of the counting function $Z(\lambda_j^{(a)}) = Q_j^{(a)}/L$ by the expression,
\begin{equation}\label{rhos}
\rho^{(a)}(\lambda^{(a)})= \frac{\mathrm{d} Z(\lambda_j^{(a)})}{\mathrm{d} \lambda_j^{(a)}},
\quad a = 1,2.
\end{equation}

In the thermodynamic $L \rightarrow \infty$ limit the Bethe equations (\ref{nbethe1},\ref{nbethe2}) turn into
coupled linear integral relations for the densities $\rho^{(a)}(\lambda^{(a)})$ which can
be solved by the Fourier transform method. The final result for the densities are,
\begin{eqnarray}\label{rhostherm}
\rho^{(1)}(\lambda^{(1)})&=& \frac{2}{\pi} \frac{\sin(\gamma) \cosh(\lambda^{(1)})}{\left [ \cosh(2\lambda^{(1)}) -\cos(2\gamma)\right ]} \nonumber \\
\rho^{(2)}(\lambda^{(2)})&=& \frac{1}{2 \pi \cosh(\lambda^{(2)})}.
\end{eqnarray}

Now from the expressions for the density $\rho^{(1)}(\lambda^{(1)})$ and Eq.\rf{spect} we
can compute the ground state energy per site $e_\infty (\gamma) = \lim_{L\to\infty} E_0 (L,\gamma)/L$.
By writing the infinite volume limit of Eq.\rf{spect} in terms of its Fourier transform we find,
\begin{equation}\label{einf}
e_{\infty}(\gamma) = -4 \sin(\gamma)  \int_{0}^{\infty} \mathrm{d} \omega \frac{
\ch[\omega(\pi/2-\gamma)] \sh[\omega(\pi-\gamma)]}{\ch[\omega \pi/2] \sh[\omega
\pi]}~~~ 
\mathrm{for}\quad 0 < \gamma \leq \frac{\pi}{2}
\end{equation}

Let us consider the behavior of the low-lying excited states about the ground state.
As usual these states are obtained from the Bethe equations (\ref{nbethe1},\ref{nbethe2}) by
making alternative choices of numbers $Q_j^{(a)}$ over the ground state configuration.
This procedure is nowadays familiar to models solved by Bethe ansatz and for technical
details see, for instance \cite{SU,EXI}. It turns out that the expressions for the
energy $\varepsilon^{(a)}(\lambda^{(a)})$ and the momenta
$p^{(a)}(\lambda^{(a)})$, measured from the ground state, of a hole excitation on
the $a$-th branch is given by
\begin{equation}\label{ene*mom}
\varepsilon^{(a)}(\lambda^{(a)}) = 2 \pi \rho^{(a)}(\lambda^{(a)}),~~~ p^{(a)}(\lambda^{(a)})= \int_{\lambda^{(a)}}^{\infty}
\varepsilon^{(a)}(x) \mathrm{d}x.
\end{equation}
To compute the dispersion relation $\varepsilon^{(a)}(p^{(a)})$ one has to eliminate the auxiliary variable
$\lambda^{(a)}$ which connects energy and momentum. This is done by first computing the integrals in Eq.\rf{ene*mom}
 with the help of the roots densities (\ref{rhostherm}). We then are able to eliminate the rapidity
 $\lambda^{(a)}$ from $\varepsilon^{(a)}(\lambda^{(a)})$ and the final results for the dispersion relations are,
\begin{eqnarray}\label{disp}
  \varepsilon^{(1)}(p^{(1)}) &=& 4 \cos(\gamma) \sin \left(\frac{p^{(1)}}{2}\right)
\sqrt{ \sin^2 \left(\frac{p^{(1)}}{2}\right)+ \tan^2(\gamma)} \nonumber \\
 \varepsilon^{(2)}(p^{(2)})  &=&  2 \sin(\gamma) \sin (p^{(2)}).
\end{eqnarray}
Note that the dispersion relation associated to particle number excitations $\varepsilon^{(1)}(p^{(1)})$
has the interesting feature of being factorized in terms of 
two physically distinct types of dispersions. In fact, the first part has a
massless behavior while the second one has a massive character with a mass term proportional
to $\tan{(\gamma)}$. By way of contrast the dispersion related to the spin branch $\varepsilon^{(2)}(p^{(2)})$
is very similar to the spin-waves of the  anti-ferromagnetic Heisenberg XXZ model. However, for low
 momenta the massless character prevails and both charge and spin excitations have a common slope at
 $p^{(a)}=0$, namely
\begin{equation}\label{disp1}
\varepsilon^{(a)}(p^{(a)}) \sim  2 \sin(\gamma) p^{(a)},
~~~\mathrm{for}~~ 0 < \gamma \leq \frac{\pi}{2}
\end{equation}
and therefore they travel with the same speed of sound $v_s = 2\sin{\gamma}$.

Let us turn our attention to the physical properties of the model at special point $\gamma = 0$.
In this case, the Hamiltonian (\ref{H}-\ref{const2}) 
commutes also with the number of local electrons pairs \cite{EKS1} and it is proportional to the graded
permutator,
\begin{equation}\label{permut}
H(\gamma=0)= \sum_{j=1}^{L} \sum_{a,b=1}^{4} (-1)^{p_a p_b} e_{ab}^{(j)} \otimes e_{ba}^{(j+1)} -L
\end{equation}
where $e_{ab}^{(j)}$ denotes 4x4 Weyl matrices acting on the $j$-th site and the Grassmann
parities are given by $p_1 = 0,\ p_2 = 1,\ p_3, = 1,$ and $p_4 = 0$.

The diagonalization of the Hamiltonian (\ref{permut}) by the Bethe ansatz was discussed in 
the literature since long ago \cite{SU,KUL}.
We remark that the respective Bethe equations do not follow immediately from Eqs.(\ref{bethe1},\ref{bethe2})
when $\gamma \to 0$ due to the peculiar pattern of the Bethe roots $\{\mu_j\}$. We find, however,
that certain properties of the  model at $ \gamma=0$ can be  inferred without the need of using its
Bethe ansatz solution. This is done by first investigating the pattern of the ground sate degeneracies of
Hamiltonian (\ref{permut}) by means of exact diagonalization up to $L=12$. This study
has reveled that the ground state sits in many different sectors whose total number of particles
is either $L$ or $L \pm 1$. This tells us the ground state for a given $L$ is $4L$-fold degenerated and that
its energy and low-lying excitations can be computed from the particular simple sectors
$N_+ = L,\ N_- = 0$ or $N_+ = 0,\ N_- = L$. Because these are typical ferromagnetic states the
calculations are rather direct. Denoting by $p$ the momentum of an excitation with spin $\sigma=-$
over the state $N_+ = L,\ N_- = 0$ one finds that the corresponding energy is,
\begin{equation}\label{disp2}
E(p) = -2L +4 \sin^2\left(\frac{p}{2}\right)
\end{equation}
where for a finite $L$ the momenta $p = \frac{2\pi}{L}K,\quad K=0,\dots,L-1$.

From Eq.\rf{disp2} we conclude that the ground state per site is
$e_\infty (\gamma=0) = -2$ and that for low momenta $p$ the excitation energy
are proportional to $p^2$. Therefore, the system has a nonrelativistic behavior
in accordance with previous works in the literature \cite{EKS1,SA}.
Interesting enough, we observe that such results can also be derived from
Eqs.(\ref{einf},\ref{disp}) by taking the limit $\gamma \to 0$. 
To obtain the ground state energy from Eq.\rf{einf} we first perform
the change of variable $\omega \to \omega/\gamma$ and afterwards take the
$\gamma \to 0$ limit. On the other hand, the dispersion relation
$\varepsilon(p) = 4\sin^2(p/2)$ follows directly from Eq.\rf{disp} by substituting
$\gamma = 0$.

We have now the basic ingredients to investigate in next section the finite-size effects
in the spectrum of the electronic model (\ref{H}-\ref{const2}) for $0 < \gamma \le \pi/2$.

\section{Critical properties}
\label{sec:crit}

The results of previous Section suggests us that the generalized Hubbard model (\ref{H}-\ref{const2}) 
in the regime $0 < \gamma \le \pi/2$ is
conformally invariant. This means that the corresponding critical properties can
be evaluated investigating the eigenspectrum finite-size corrections \cite{CAR}.
For periodic boundary conditions, the ground state $E_0 (L,\gamma)$ are expected to
scale as,
\begin{equation}\label{cc}
\frac{E_0(L,\gamma)}{L} =e_{\infty} -\frac{\pi v_s(\gamma) c}{ 6 L^2}
+ O\left( L^{-2} \right),
\end{equation}
where $c$ is the central charge.

From the excited states $E_{\alpha}(L,\gamma)$ we are able to determine the
dimensions $X_{\alpha}(\gamma)$ of the respective primary operators, namely
\begin{equation}\label{tower}
\frac{E_{\alpha}(L,\gamma)}{L}
-\frac{E_{0}(L,\gamma)}{L}
=\frac{2 \pi v_s(\gamma) X_{\alpha}(\gamma)}{ L^2}
+ O\left( L^{-2} \right).
\end{equation}

A first insight on the structure of the finite-size corrections can be obtained by
applying the so-called density root method \cite{RO,RO1,RO2}. This approach explores
the Bethe ansatz solution and it makes possible to compute the $O\left( L^{-2} \right)$
corrections to the densities of roots $\rho^{(a)}(\lambda^{(a)})$. This method is however
only suitable for systems whose ground state and low-lying excitations are described by
real roots. Fortunately, this is exactly the situation we have found in Section \ref{sec:thermod}
once the second root is shifted by $i\pi/2$. Considering this subtlety on the root density approach
we find that the leading finite-size  behavior of the eigenenergies is,
\begin{equation}\label{FSC}
\frac{E(L,\gamma)}{L}=  e_{\infty} (\gamma) +
\frac{2\pi}{L^2} v_{s}(\gamma) \left[ -\frac{1}{6} + X_{n_{+},n_{-}}^{m,m_{-}}(\gamma) \right]
+ O\left( L^{-2} \right),
\end{equation}
where the dependence of the scaling dimensions
$\displaystyle X_{n_{+},n_{-}}^{m,m_{-}}(\gamma)$ on the anisotropy $\gamma$ is,
%
%\begin{widetext}
\begin{equation}\label{dimens}
X_{n_{+},n_{-}}^{m,m_{-}}(\gamma)= \frac{1}{4} \left[ n_{+}^2 + n_{-}^2 +2(1-\frac{2\gamma}{\pi}) n_{+} n_{-} \right] +
\frac{\pi^2}{4 \gamma (\pi-\gamma)} \left [m^2 + m_{-}^2 -2(1-\frac{2\gamma}{\pi}) m m_{-} \right].
\end{equation}
%\end{widetext}
%
As before the integers $n_\pm$ parameterizes the numbers of electrons $N_\pm = L/2 - n_\pm$ with spin $\sigma = \pm$.
The indices $m = m_+ + m_-$ and $m_+$ characterize the presence of holes in the $Q_j^{(1)}$ and 
$Q_j^{(2)}$ distributions
and in principle can be integers or half-integers. This approach is however not able to predict either
the possible values for the vortex numbers $m$ and $m_+$ as well possible constraints with the 
corresponding spin-wave
integers $n$ and $n_+$.
To shed some light on this problem we shall first study the finite-size effects at the
particular point $\gamma = \pi/2$. For $\gamma = \pi/2$ we see that all the interactions
in the Hamiltonian (\ref{H}-\ref{const2}) cancel out and we remain with
two coupled free fermion models. In this case standard Fourier technique is able
to provide us the exact expressions for the low-lying energies in the case of arbitrary $L$.
The respective calculations depend on the total number of electrons on the lattice $L$.
We find that when $n = n_+ + n_-$ is odd that the expression for the lowest energy in
this sector is given by,
\begin{equation}\label{odd}
E_{\mathrm{odd}}(L,\frac{\pi}{2}) = - 2
\frac{\left[ \cos{\left(\frac{\pi n_{+}}{L} \right)} + \cos{\left(\frac{\pi n_{-}}{L} \right)} \right]}{\sin{\left( \frac{\pi}{L} \right)}}.
\end{equation}
Considering the asymptotic expansion of Eq.\rf{odd} for large $L$ one finds,
\begin{equation}\label{oddbulk}
\frac{E_{\mathrm{odd}}(L,\frac{\pi}{2})}{L} = e_{\infty} (\frac{\pi}{2}) +
\frac{2\pi}{L^2} v_{s}(\frac{\pi}{2}) \left[ -\frac{1}{6} + \frac{n_{+}^2 + n_{-}^2}{4} \right]
+ O\left( L^{-2} \right).
\end{equation}
By way of contrast when $n = n_+ + n_-$ is an even number the lowest energy is,
\begin{equation}\label{even}
E_{\mathrm{even}}(L,\frac{\pi}{2}) = -\sum_{\sigma=\pm}
\frac{\cos \left[\frac{\pi (n_{\sigma}+1)}{L} \right]}  
{\sin \left( \frac{\pi}{L} \right)}+
\frac{\cos \left[\frac{\pi (n_{\sigma}-1)}{L} \right]} 
{\sin \left( \frac{\pi}{L} \right)}
\end{equation}
whose expansion for large $L$ is,
\begin{eqnarray}\label{evenbulk}
\frac{E_{\mathrm{even}}(L,\frac{\pi}{2})}{L} &=& e_{\infty} (\frac{\pi}{2}) +
\frac{2\pi}{L^2} v_{s}(\frac{\pi}{2}) \left[ -\frac{1}{6} + \frac{n_{+}^2 + n_{-}^2}{4} + \frac{1}{2} \right] \nonumber \\
& & + O\left( L^{-2} \right).
\end{eqnarray}

Taking into account Eqs.(\ref{oddbulk},\ref{evenbulk}) we see that the expected finite size corrections depend
whether the index $n$ is an odd or even integer. In addition, by comparing Eqs.(\ref{oddbulk},\ref{evenbulk})
with the general results (\ref{FSC},\ref{dimens}) at $\gamma = \pi/2$ we clearly see that for $n$ odd
the numbers $m$ and $m_-$ appear to start from zero while for $n$ even the lowest allowed
value for $m$ and $m_-$ is in fact one-half. This analysis strongly suggests that possible values
for the vortex numbers $m$ and $m_+$ should satisfy the following rule
\begin{eqnarray}\label{const3}
\bullet \;\;\; \mathrm{for} \;\; n \;\; \mathrm{odd}  \;\;\; &&\rightarrow \;\;\;\;\;  m,m_{-} = 0, \pm 1 ,\pm 2, \dots \nonumber \\
\bullet \;\;\; \mathrm{for} \;\; n \;\; \mathrm{even} \;\; &&\rightarrow \;\;\;\;\;
m,m_{-} = \pm \frac{1}{2}, \pm \frac{3}{2} ,\pm \frac{5}{2} , \dots \quad .
\end{eqnarray}

Let us now check if the above proposal remains valid for other values of the
parameter $\gamma$. This is done mostly by solving numerically the original Bethe
equations (\ref{bethe1}, \ref{bethe2}) up to $L=32$. For the excited states whose
respective Bethe roots are unstable already for moderate values of $L$ we have
used the data obtained from the numerical diagonalization through the Lanczos method.
This numerical work enables us to compute for each $L$ the following sequence
\begin{equation}\label{FSSestim}
X(L) = \left( \frac{E(L,\gamma)}{L} - e_{\infty}(\gamma) \right) \frac{L^2}{2\pi v_{s}(\gamma)} +
\frac{1}{6}
\end{equation}
\begin{samepage}
By extrapolating $X(L)$ for several values of $L$ we are able to verify the expression (\ref{dimens}) for
$X_{n_{+},n_{-}}^{m,m_{-}}(\gamma)$ and the constraints (\ref{const3}).
In Tables \ref{one}, \ref{two}, and \ref{three} we exhibit the finite-size sequence \rf{FSSestim}
for six lowest dimensions on the even sector to make an extensive check of the less unusual
part of the rule (\ref{const3}). For sake of completeness we also present
three conformal dimensions corresponding to the $n$ odd sector.  All those numerical results
confirm the conjecture (\ref{dimens}, \ref{const3}) for the finite size properties
of the generalized Hubbard model (\ref{H}-\ref{const2}).

%
%\begingroup
%\squeezetable
\begin{table*}
\begin{tabular}{|c||c|c|c|c|c|c|}
  \hline
  $L$ & $X_{0,0}^{\half,\half}(\frac{\pi}{5})$ & $X_{0,0}^{\half,-\half}(\frac{\pi}{5})$ & $X_{1,-1}^{\half,\half}(\frac{\pi}{5})$ & $X_{0,0}^{\half,\half}(\frac{\pi}{3})$ & $X_{0,0}^{\half,-\half}(\frac{\pi}{3})$ & $X_{1,-1}^{\half,\half}(\frac{\pi}{3})$ \\ \hline \hline
  8      & 0.313380 & 1.227954 & 0.488672 & 0.380231 & 0.752958 & 0.681073 \\ \hline
  12     & 0.312980 & 1.239120 & 0.498637 & 0.377310 & 0.751250 & 0.694030 \\ \hline
  16     & 0.312787 & 1.243526 & 0.502997 & 0.376297 & 0.750687 & 0.699292 \\ \hline
  20     & 0.312689 & 1.245526 & 0.505395 & 0.375829 & 0.750434 & 0.701996 \\ \hline
  24     & 0.312633 & 1.247012 & 0.506892 & 0.375575 & 0.750299 & 0.703591 \\ \hline
  28     & 0.312595 & 1.247954  & 0.507906 & 0.375423 & 0.750218 & 0.704619 \\ \hline
  32     & 0.312576 & 1.248063  & 0.508633 & 0.375323 & 0.750166 & 0.705327 \\ \hline \hline
  Extrap.& 0.31249$\pm 1$ & 1.2504$\pm 2$ & 0.51219$\pm 1$ & 0.37498$\pm 1$ & 0.74999$\pm 1$ & 0.708336$\pm 1$ \\ \hline \hline
  Exact  & 0.3125 & 1.25 & 0.5125 & 0.375 & 0.75 & 0.70833$\dots$ \\ \hline
\end{tabular}
\caption{Finite size sequences \rf{FSSestim} of the anomalous dimensions for $\gamma = \pi/5,\ \pi/3$ from the Bethe
ansatz. The expected exact conformal dimensions are $X_{0,0}^{\half,\half}(\gamma)=\frac{1}{4(1-\gamma/\pi)}$,
 $X_{0,0}^{\half,-\half}(\gamma)=\frac{1}{4(\gamma/\pi)}$, $X_{1,-1}^{\half,\half}(\gamma)=\frac{\gamma}{\pi}+
 \frac{1}{4(1-\gamma/\pi)}$.
 }\label{one}
\end{table*}
%\endgroup
%
%\begingroup
%\squeezetable
%
\begin{table*}
\begin{tabular}{|c||c|c|c|c|c|c|}
  \hline
  $L$ & $X_{1,0}^{0,0}(\frac{\pi}{5})$ & $X_{2,-1}^{0,0}(\frac{\pi}{5})$ & $X_{2,-2}^{\half,\half}(\frac{\pi}{5})$ &
  $X_{1,0}^{0,0}(\frac{\pi}{3})$ & $X_{2,-1}^{0,0}(\frac{\pi}{3})$ & $X_{2,-2}^{\half,\half}(\frac{\pi}{3})$ \\ \hline \hline
  8      & 0.251098 & 0.642630 & 1.000395 & 0.252587  & 0.902529 & 1.541194 \\ \hline
  12     & 0.250523 & 0.646574 & 1.047807 & 0.251149 & 0.910140 & 1.622978 \\ \hline
  16     & 0.250301 & 0.648003 & 1.068757 & 0.250646 & 0.912924 & 1.655748  \\ \hline
  20     & 0.250195 & 0.648689 & 1.080196 & 0.250413 & 0.914244 & 1.672266  \\ \hline
  24     & 0.250136 & 0.649072 & 1.087265 & 0.250287 & 0.914971 & 1.681827   \\ \hline
  28     & 0.250101 & 0.649307 & 1.092001 & 0.250211 & 0.915414 & 1.687896   \\ \hline
  32     & 0.250077  & 0.649463 &1.095375 & 0.250161 & 0.915704 & 1.692009 \\ \hline \hline
  Extrap.& 0.250003$\pm 1$ & 0.65003$\pm 1$ & 1.1124$\pm 1$ & 0.250004$\pm 2$ & 0.9167$\pm 2$ & 1.70825$\pm 1$ \\ \hline \hline
  Exact  & 0.25 & 0.65 & 1.1125 & 0.25 & 0.91666$\dots$ & 1.70833$\dots$ \\ \hline
\end{tabular}
\caption{Finite size sequences \rf{FSSestim} of the anomalous dimensions for $\gamma = \pi/5,\ \pi/3$ from the Bethe
ansatz. The exact conformal dimensions are $X_{1,0}^{0,0}(\gamma)=\frac{1}{4}$,
 $X_{2,-1}^{0,0}(\gamma)=\frac{1}{4}+\frac{2\gamma}{\pi}$, $X_{2,-2}^{\half,\half}(\gamma)=\frac{4\gamma}{\pi}+
 \frac{1}{4(1-\gamma/\pi)}$.
 }\label{two}
\end{table*}
%\endgroup
%
%\begingroup
%\squeezetable
\begin{table*}
\begin{tabular}{|c||c|c|c|c|c|c|}
  \hline
  $L$ & $X_{1,0}^{1,0}(\frac{\pi}{5})$ & $X_{1,1}^{\half,\half}(\frac{\pi}{5})$ & $X_{1,1}^{\half,-\half}(\frac{\pi}{5})$ & $X_{1,0}^{1,0}(\frac{\pi}{4})$ & $X_{1,1}^{\half,\half}(\frac{\pi}{4})$ & $X_{1,1}^{\half,-\half}(\frac{\pi}{4})$ \\ \hline \hline
  4      & 1.497964 & 1.395242 & 1.426027 & 1.355168 & 1.225412 & 1.261369 \\ \hline
  6      & 1.661634 & 1.288542 & 1.741570 & 1.480050 & 1.172763  & 1.563850 \\ \hline
  8      & 1.724314 & 1.224764 & 1.888707 & 1.523739 & 1.140033 & 1.639713 \\ \hline
  10     & 1.754379 & 1.188829 & 1.947347 & 1.544502 & 1.121710 & 1.679436 \\ \hline
  12     & 1.771238 & 1.167328 & 1.977593 & 1.556018 & 1.110818 & 1.699900 \\ \hline
  14     & 1.781660 & 1.153629 & 1.995956 & 1.563070 & 1.103911 & 1.712660 \\ \hline
  16     & 1.788561 & 1.144426 & 2.008012 & 1.567703 & 1.098013  & 1.721046 \\ \hline \hline
  Extrap.& 1.812$\pm 1$ & 1.12$\pm 1$ & 2.06$\pm 1$ & 1.583$\pm 1$ & 1.079$\pm 1$ & 1.73$\pm 1$ \\ \hline \hline
  Exact  & 1.8125 & 1.1125 & 2.05 & 1.5833$\dots$ & 1.0833$\dots$ & 1.75 \\ \hline
\end{tabular}
\caption{Finite size sequences \rf{FSSestim} of the anomalous dimensions for $\gamma = \pi/5,\ \pi/4$ from Lanczos.
The expected exact conformal dimensions are $X_{1,0}^{1,0}(\gamma)=\frac{1}{4}+\frac{1}{4(\gamma/\pi)(1-\gamma/\pi)}$,
 $X_{1,1}^{\half,\half}(\gamma)=(1-\gamma/\pi)+\frac{1}{4(1-\gamma/\pi)}$, $X_{1,1}^{\half,-\half}(\gamma)=
 (1-\gamma/\pi)+\frac{1}{4(\gamma/\pi)}$.
 }\label{three}
\end{table*}
%\endgroup
%
%\begingroup
%\squeezetable
%
\end{samepage}

We shall now proceed with a discussion of the results obtained so far. From Section \ref{sec:thermod} we know
that the ground state sits in the sectors $n_+ = \pm 1$ and $n_- = 0$ or $n_+ = 0$ and $n_- = \pm 1$.
Considering the rule \rf{const3} the corresponding vortex numbers have the lowest possible values
 $m=m_+ =0$ and from Eqs.(\ref{FSC},\ref{dimens}) we derive the following finite size behavior,
\begin{equation}\label{E0}
\frac{E_0(L,\gamma)}{L} = e_{\infty} +\frac{\pi v_s(\gamma)}{ 6 L^2}
+ O\left( L^{-2} \right).
\end{equation}

Direct comparison between Eq.(\ref{cc}) and Eq.(\ref{E0}) 
leads us to conclude that the central charge of 
the underlying conformal theory is,
\begin{equation}\label{cc2}
c=-1~~~~
\mathrm{for}~~~ 0< \gamma \leq \frac{\pi}{2}
\end{equation}

The conformal dimensions of the primary operators $\bar{X}_{n,n_+}^{m,m_+}(\gamma)$ depend
on the anisotropy $\gamma$ and they should be measured from the ground state
$E_0 (L,\gamma)$. Considering Eqs.(\ref{FSC},\ref{dimens}) together
with Eq.\rf{E0} we find that they are given by,
\begin{equation}\label{dim2}
{\bar X}_{n,n_{+}}^{m,m_+}(\gamma) =
X_{n,n_{+}}^{m,m_{+}}(\gamma) -
\frac{1}{4}
~~~~\mathrm{for}~~~ 0< \gamma \leq \frac{\pi}{2}
\end{equation}

To our knowlodge, models exhibiting this kind of
universality class have so far been found in a not self-adjoint
theory  based on the deformed $osp(2|2)$ symmetry \cite{GP}. Therefore, the correlated 
electron system (\ref{H}-\ref{const2})  
appears to be the first example of a Hermitian Hamiltonian
whose continuum limit is described by a field theory with $c=-1$ with
continuously varying anomalous dimensions.  The fact that a line of critical
exponents with $c<0$ can be realized in terms of Hermitian models could be
of importance for  practical applications in condensed matter such as in the physics of
disordered systems. 

\section{Conclusions}

We have studied the physical properties of an exactly solvable generalization
of the Hubbard model with free parameter $\gamma$ related to the quantum
$U_q[SU(2|2)$ superalgebra where $q=\exp(i \gamma)$. We have determined the
nature of the ground state and the  behavior of the elementary
excitations. The peculiar feature of the model is that the dispersion
relation for the charge sector is given in terms of the product of 
massless and massive energy-momenta relations. In the regime $ 0 < \gamma \le \pi/2$
the low-lying excitations have a relativistic behavior and the underlying
conformal theory has central charge $c=-1$ and a line of continuously
varying exponents. For the particular point $q=1$ we have
argued that basic properties can be obtained without recourse to
the Bethe ansatz solution. We expect that this observation remains
valid for all integrable models based on the Lie superalgebra
$sl(p|q)$ for arbitrary finite number of bosonic ($p$) and
fermionic ($q$) degrees of freedom.  This suggests that
the models based on the
deformed $sl(p|q)$ symmetry may also have excitation modes with dispersion
relation exhibiting both massless and massive behaviors. We hope to investigate
this interesting possibility as well as its consequences in a future publication.

\section{Acknowledgments}

This work has been supported by the Brazilian Research Agencies CNPq and FAPESP.

\end{document}